# Analysing the Assumed Benefits of Software Requirements

Richard Ellis-Braithwaite

Loughborough University, United Kingdom
r.d.j.ellis-braithwaite@lboro.ac.uk

**Abstract.** Often during the requirements engineering (RE) process, the value of a requirement is assessed, e.g., in requirement prioritisation, release planning, and trade-off analysis. In order to support these activities, this research evaluates Goal Oriented Requirements Engineering (GORE) methods for the description of a requirement's value. Specifically, we investigate the goal-to-goal contribution relationship for its ability to demonstrate the value of a requirement, and propose that it is enriched with concepts such as correlation, confidence, and utility.

## 1 Problem Statement

There is often a "field-of-dreams" assumption that once software is built to the specified requirements, benefit will come [1]. The fact that there is little correlation between a company's level of IT investment and its profitability or market value, leading to the so-called "information paradox" [2], highlights the dangers of this assumption. Therefore, stakeholders responsible for a software project's funding need to be able to demonstrate that the software will be beneficial. Furthermore, practitioners performing RE processes where the benefit of a requirement is assessed (e.g., in prioritisation) need to know how benefit is defined by the stakeholders, and then how the requirements will contribute to it. With this in mind, this research aims to explore, improve, and evaluate methodologies for analysing the assumed benefits of requirements, and thereby, the alignment of those requirements to business strategy. Through evaluation of the methods in industrial projects, we aim to optimise the cost to benefit ratio of their application through methodology improvements, guidelines, and tool support.

The following research questions were formulated in response to problems faced by our industry partners, with the overall goal to improve decision-making in the RE process via better communication and analysis of assumed benefit:

- RQ1. What evidence exists to show that implemented requirements (i.e., software features and qualities) are not always beneficial?
- RQ2. What is an appropriate approach for modelling the assumed benefits of software requirements?
- RQ3. What aspects of the resulting benefit model are important for analysing the strategic alignment of software requirements?
- RQ4. What are the quality characteristics of such models, and what challenges preclude them?
- RQ5. How can a supporting tool address the challenges elicited from RQ4?

## 2  Motivation

The oft-cited, but ageing CHAOS report [3] suggests that two thirds of functional requirements which are specified before implementation are never or rarely used after their implementation. This claim is supported by independent surveys, for example [4], which found that only 27% of the functionality in word processing software is ever used. Since stakeholders are often "motivate[d] to brainstorm requirements which they think that they just might need at some point" [5], it should be no surprise that some requirements lack pertinence, and as a result, deplete development resources.

Pertinence also affects the quality of non-functional requirements (NFR's). For example, a reliability requirement stipulating a certain level of service uptime should be the result of a trade-off made between two or more conflicting stakeholder goals, e.g., "maximize service availability" and "minimize infrastructure costs". These trade-offs aim to maximise the utility of the software by optimising the associated cost-benefit ratios and acceptable risk levels [6]. Unfortunately, it has been found that stakeholders are rarely the source of NFR's, since "architects consider themselves to be the real experts when it comes to defining efficiency, reliability, and other similar aspects" [7]. On the contrary, the RE activity is primarily concerned with the description of the application domain [8] - the stakeholder's specialism, rather than the machine, which is the architect's specialism. In order to perform a successful trade-off, the rationale behind each goal is required, and without stakeholder involvement, over/under specification of NFR's will likely occur as a result of developer assumptions, ultimately leading to increased costs, delays, or in extreme cases, project failure [9]. Thus, for such consequences to be avoided, developers and stakeholders need to be able to comprehend the effects of a requirement's implementation on each other's goals, since the quality of any decision is underpinned by the information available to support it [10].

Numerous surveys blame the majority of software project failures, including poor return on investment (ROI), on inadequate RE [3, 11] - or more specifically, on poor stakeholder communication and incorrect assumptions [12]. Framing the problem in the context of these failure factors, we wish to minimise assumptions made about the benefits that stakeholders expect, by communicating those expectations to developers. Regardless of the cause, as a result of IT's poor ROI, IT-business strategy alignment has been the top ranking concern of business executives for the last two decades [13].

## 3  Related Work

The value based software engineering (VBSE) agenda [14] is motivated by the observation that most software projects fail because they don't deliver stakeholder value, yet, much software engineering practice is done in a value-neutral setting (e.g., where project cost and schedule is tracked rather than stakeholder value). Value-based requirements engineering (VBRE) takes the economic value of IT products into perspective through stakeholder identification, business case analysis, requirements prioritisation, and negotiation [15]. The primary VBRE methods are Business Case Analysis (BCA) and Benefits Realization Analysis (BRA) [16]. We consider other VBRE processes (e.g., prioritisation) as secondary, since they depend on benefit estimation.

In its simplest form, BCA involves calculating a system's ROI (financial benefits versus costs, in present value). An advancement from BCA, $e^3$value modelling seeks to understand the economic value of a system by mapping value exchanges between actors, ultimately leading to financial analysis such as discounted cash flow [17]. Howev-er, these approaches are complex in their application, since the validity of any concise financial figure depends on assumptions holding true, e.g., that independent variables remain within expected ranges. Estimating benefit involves further intricacies such as uncertainty, and the translation of qualitative variables (e.g., software user happiness) to quantitative benefits (e.g., sales revenue) - none of which are made explicit by BCA.

BRA's fundamental concept is the Results Chain [18], which visually demonstrates traceability between an initiative (i.e., a software system) and its outcomes (i.e., benefits) using a directed graph, where nodes represent initiatives, outcomes, and assumptions, and edges represent contribution links. BRA's contribution links allow one initiative to achieve multiple outcomes, but the links are not quantitative, e.g., outcome: "reduced time to deliver product" can contribute to outcome: "increased sales" if assumption: "delivery time is an important buying criterion" holds true – but the quantitative relationship between "product delivery time" and "sales increase" is not explored. This poses a problem when outcomes are business objectives, since the satisfaction of an objective depends on the extent that it is contributed to, e.g., in the case of a cost reduction objective, the extent is the amount of reduction that will be contributed [19].

Goal Oriented Requirements Engineering (GORE) methods are capable of demonstrating alignment between software requirements and goals with AND/OR goal graphs [20]. Goal graphs ensure the pertinence of software requirements [20], since requirements must trace to a more abstract goal to explain the rationale for the requirement's implementation. Additionally, goal graphs improve communication between stakeholders since requirements are restated at various levels of abstraction (through goals), thereby bridging the gap between technically minded developers and application-domain focused stakeholders. Quantified goal graphs are used in [20] to model how far one goal contributes to its parent goals. However, the approach does not translate contribution scores toward various levels of goal abstraction, and therefore does not place the benefit contributed by a requirement into context, e.g., that a large saving may only be derived from a small cost [19]. A probabilistic layer for quantified goal graphs is proposed in [21] to reason about the probability that a goal's more abstract parent goals will be satisfied. However, this approach is time consuming, has limited applications, and does not capture stakeholder "attitude, preference and likings" [22].

Singh and Woo review the IT-business alignment literature within the RE field [23], and conclude that the majority of frameworks do not address business strategy or value analysis. The Strategic Alignment Model [24] is proposed as a theoretical framework for conceptualising IT-business alignment, but it is not taken beyond the conceptual level, and thus does not consider traceability to system requirements [25]. The OMG's standardised Business Motivation Model (BMM) [26] states that a set of quantified business objectives forms a business strategy, and that satisfaction of the objectives satisfies the strategy. Several methodologies have used the BMM to show the strategic alignment of software requirements. For example, in [27], OMG's SysML requirements metamodel is extended to establish traceability to the "tactic" concept from the BMM. Similarly, in [25], the BMM is used to decompose business strategy down to software requirements (i.e., from vision statements to tasks) which will satisfice the strategy. However, much like the BRA, neither approach takes a quantitative approach to demonstrating strategic alignment, yet strategic alignment depends on the extent of an objective's satisfaction, which is measured quantitatively.

## 4 Proposed Solution

The foundations for our approach to answering RQ2 are introduced in [19]. A brief summary is that the benefits of software requirements are stated at various levels of abstraction using goal graphs (where a benefit is the advantage gained by treating a problem and where goals are inverted problems). Goal graphs are used because they are well suited for visualising abstraction and refinement (hierarchies between parent/child goals), and can also visualise dependencies and cause-effect relationships [28].

A more detailed summary is that GRL goal graphs [29] are used to relate software requirements (represented by GRL task elements) to business objectives (represented by GRL hard goal elements) through GRL contribution links. Software requirements are represented by GRL tasks, in that it is the task of implementing the requirement that should satisfy some business objective. Business objectives are specified using the GQM+Strategies template [30], such that objectives can be considered as GRL hard goals (i.e., the objective's required magnitude is set). The proposed approach is: a) *prescriptive* in that some amount of business objective satisfaction is prescribed, b) *descriptive* in that the problem to be solved by the objectives is contextualised by various levels of goal abstraction, and c) *predictive* in that the contributions made by requirements or objectives to higher-level business objectives are quantitatively estimated.

In order to illustrate the approach's application, we refer to a software project that the authors were involved with. The software manages and schedules media files for digital signage (advertising). Fig. 1 shows an excerpt of a goal graph in the context of this project (in accordance with [19]), which explores some assumed benefits of a proposed functional requirement. A reader unfamiliar with GRL can find guidance in [29].

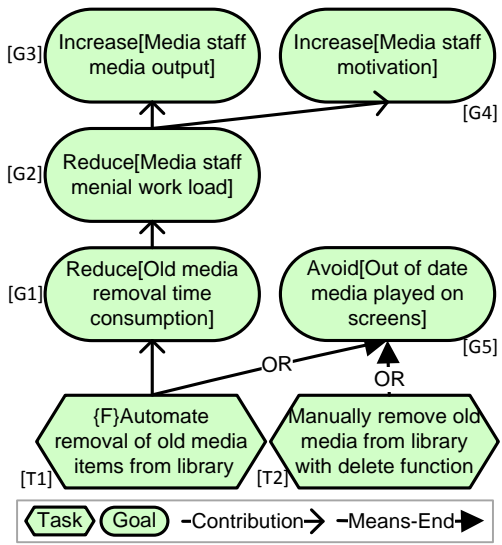

**Fig. 1.** Example goal graph showing the abstracted benefits of a proposed functional requirement

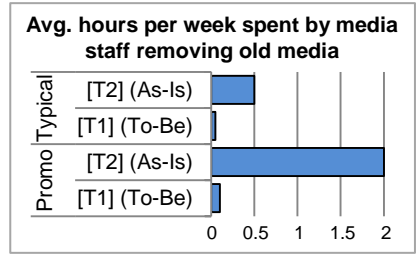

**Fig. 2.** Contribution of [T1] → [G1]

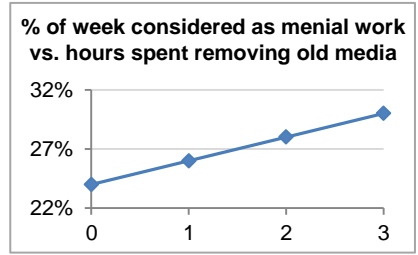

**Fig. 3.** Contribution of [G1]→[G2]

When it comes to answering RQ3, we are most interested in the contribution links between software requirements and business objectives, since they represent the alignment. Specifically, we are interested in describing *GoalX* to *GoalY* contribution such that various levels of satisfaction of *GoalX* are mapped to various levels of satisfaction of *GoalY*. We are most interested in causal relationships between *GoalX* and *GoalY,* but in reality correlation and causation are difficult to distinguish in sociotechnical systems [31]. We also take into account the effects of the system's various usage profiles on goal contribution, as shown in Fig. 2 by the y-axis grouping (the "Promo" group represents a promotional period, such as a seasonal holiday).

For functional requirements, the contribution data is discrete, since there are only two states to capture - when the requirement is, or is not implemented. Thus, in Fig. 2, we refer to [T2] as the As-Is state to represent when the requirement is not implemented, and [T1] as the To-Be state to represent when the requirement is implemented. For non-functional requirements, the contribution data is continuous, since when properly specified, their values are numerical (e.g., an uptime requirement will have various possible states of satisfaction: 99.99%, 99.98%, etc.). Thus, if [T1] were non-functional, e.g., "the maximum time an expired media item should be displayed for is 5 minutes", then Fig. 2 would be an XY graph rather than a bar graph, with x-axis values specified in minutes and y-axis values derived from the goal that it contributes to. Note that soft goals are not used, since business objectives are defined quantitatively according to the BMM, and thus, qualitative contribution to quantitative goals is ambiguous.

Once the first contribution link has been described, we move up the goal graph to describe the next contribution link, as in Fig. 3, to provide context to the benefits (in Fig. 3, we show that the menial work is only partly comprised of old media removal). Enriching the contribution links with this information will allow "what-if?" analysis through interpolation and contribution propagation. Additional benefits include better elicitation of goal tolerance levels and improved clarification of the goal's criticality, i.e., the extent to which a goal's satisfaction criteria can be stretched (e.g., relaxing a non-functional requirement) without causing failure of the goals it that it contributes to.

To enrich the contribution link further, we make confidence explicit, that is, when a practitioner describes the correlation between two goals, their confidence in their descriptions will vary depending on their expertise and previous experience. Thus, capturing confidence (e.g., between 0 and 1) for each data point in a contribution link will allow decisions to be made in consideration of the assigner's uncertainty (lack of knowledge), previous accuracy in their confidence assignments, and their risk preference. Alternative approaches to representing confidence will be evaluated (e.g., enumerated estimate points {worst-case, likely and best-case}, intervals, and probability distributions).

Finally, root goals (those which do not contribute to other goals) are mapped to utility [10], where various levels of goal satisfaction result in various levels of "goodness". For example, referring to the root goal [G4] in Fig. 1, the various levels of staff motivation (e.g., measured on a likert scale between 0 and 5) would map to utility values (e.g., between 0 and 1). This will allow non-linear relationships between motivation and the utility of that motivation to be represented (e.g., that the difference between 0 and 1 on the motivation scale is bigger than 4 and 5). The concept of utility is both subjective and specific to the utility assigner. However, capturing it will explain the criticality of a root goal's satisfaction criteria, and any differences in utility assignment between stakeholders will be made apparent before the requirement is implemented. Thus, for this to be useful, the stakeholder's utility preferences must be communicated to each other for conflict resolution. Then, the stakeholder's utility functions can be aggregated in order to improve their integrity, as in the "wisdom of the crowd" theory [32].

RQ4 examines the qualities of a model used for the analysis of strategic alignment. The elicited qualities are much like the "completeness" quality of a requirements doc-ument in that they are aspirational (their complete satisfaction is not expected). So far, we have elicited three core qualities from the application of our approach:

- **Determinism** – a goal graph produced by one practitioner relating a requirement to an organisation's goals should be similar to one produced by another. This can be supported by defining goals with metrics formalisation templates.
- **Transparency** – a goal graph should be self-explanatory rather than reliant on assumed widespread knowledge of what a goal means or why it matters. This can be achieved with goal abstraction to place goals into context.
- **Reusability** – the applicability of goal definitions (and their associated contribution links) to future projects should be as wide as possible. An example of poor reusability would be describing a contribution link relatively (e.g., "10% user task time reduction leads to 20% task cost reduction") without providing the absolute figures (e.g., $x$ dollars were saved by reducing $y$ hours).

The primary challenge to the first two qualities is that objective data is far rarer than subjective data. Decision makers have found favour with inferior processes (e.g., qualitative goal contribution scoring) because they do not force you to think very hard [10]. For example, most managers will have the opinion that reducing menial work will improve employee motivation. Thus, assigning a "+" to that contribution link is easy, but understanding the extent to which menial work affects motivation requires a survey, since it might be the case that employees are unaffected by, or even enjoy menial work. Future work will attempt to understand which goal contribution links are in most need of analysis, since it is recognised that practitioner time is finite.

RQ5 looks at how far a tool can support the application of the proposed approach. A tool is currently under development[1], whose current features were derived from inefficiencies encountered whilst implementing the approach. For example, automatic goal graph drawing resulted from the observation that drawing large goal graphs is time consuming due to the number of edge collisions that occur. We plan to extend the tool's functionality with features such as goal similarity analysis (duplicate detection) in order to support the reuse of data from previous projects (e.g., benefits of a feature).

## 5 Research Method, Progress & Novelty

This PhD project is in collaboration with two industrial partners (LSC Group and Rolls-Royce)[2]. One industrial partner will provide case studies for software developed for their own organisation, while the other will provide case studies for software developed for an external organisation. The research approach adopted for this thesis is based on the experimental software engineering paradigm [33]. Firstly, a problem was identified with the help of our industrial partners. Structured interviews and questionnaires were then used to investigate the problem, which complements the motivation identified from the literature (briefly discussed in Section 2). Then, the scientific problem was defined in the format of research questions. After that, a solution idea was formed following a systematic literature review. A prototype tool was then developed to make the required data's capture, representation, and analysis possible so that the solution idea can be improved through feedback. The solution idea will then be validated against the scientific and the practical problem using case studies for evaluation.

---

[1] The prototype tool is available to download at http://www.goalviz.info/REFSQ-DS/
[2] The author wishes to thank Dr. Tim King, Dr. Badr Haque & Ralph Boyce for their support as industrial supervisors, and Dr. Russell Lock and Prof. Ray Dawson as academic supervisors.

This PhD project is half way through its three year duration, and an initial solution has been outlined. The remainder of the project will focus on case study research to evaluate the effectiveness of the approach compared to the current state of the art. Initial feedback from the tool's evaluation with practitioners is promising. Stakeholders, especially business managers, are attracted to the ability to understand technical software requirements in terms of the business objectives that they are familiar with. Feedback from developers has been more critical, since they are evaluated on the quality and timeliness of their programming, rather than on the value aspects of the software. Additionally, the numerical aspect of the approach has been off-putting to some.

As for novelty, we are not aware of an approach that considers the benefits of a software requirement as a chain of quantified goal abstractions. In particular, we are not aware of an approach that attempts to forecast the effect of a goal's satisfaction on its parent goal(s) at:
   a) varying levels of goal satisfaction extent (explaining the effects of partial/full requirement satisfaction on multiple levels of goal abstraction);
   b) varying levels of software usage (explaining the different profiles of software usage that can affect a requirement's contribution to goals);
   c) varying levels of stakeholder confidence (explaining the extent to which a requirement's satisfaction may not contribute to a goal as specified);
   d) varying levels of stakeholder utility (explaining the non-linear relationships between the extent of a goal's satisfaction and the utility gained);
   e) varying levels of stakeholder agreement (explaining the variance between the stakeholder's estimates about the benefits that will be realised).